\documentclass[prd,preprint]{revtex4-1} 
\usepackage[pdftex]{}
\usepackage{graphicx}
\usepackage{latexsym}
\usepackage{amssymb}
\usepackage{bm}

\newcommand{\f}{\begin{equation}}
\newcommand{\ff}{\end{equation}}

\newcommand{\cg}{\ensuremath{\tilde{\gamma}}}
\newcommand{\ta}{\ensuremath{\tilde{A}}}

\newcommand{\tg}{\ensuremath{\tilde{\Gamma}}}
\newcommand{\tR}{\ensuremath{\tilde{R}}}

\newcommand{\beq}{\begin{equation}}
\newcommand{\eeq}{\end{equation}}
\newcommand{\bea}{\begin{eqnarray}}
\newcommand{\eea}{\end{eqnarray}}

  \def\msol{\ifmmode M_\odot\else$M_\odot$\fi}
  
\begin{document}


\title{Extracting Gravitational Waves Induced by Plasma Turbulence in the Early Universe through an Averaging Process}

\author{David Garrison}
\address{Physics Department, University of Houston Clear Lake, Houston, Texas 77058}

\author{Christopher Ramirez}
\address{Physics Department, University of Houston Clear Lake, Houston, Texas 77058}


\begin{abstract}
This work is a follow-up to the paper, "Numerical Relativity as a Tool for Studying the Early Universe".  In this article, we determine if cosmological gravitational waves can be accurately extracted from a dynamical spacetime using an averaging process as opposed to conventional methods of gravitational wave extraction using a complex Weyl scalar.  We calculate the normalized energy density, strain and degree of polarization of gravitational waves produced by a simulated turbulent plasma similar to what was believed to have existed shortly after the electroweak scale.  This calculation is completed using two numerical codes, one which utilizes full General Relativity calculations based on modified BSSN equations while the other utilizes a linearized approximation of General Relativity.  Our results show that the spectrum of gravitational waves calculated from the nonlinear code using an averaging process are nearly indistinguishable from those calculated from the linear code.  This result validates the use of the averaging process for gravitational wave extraction of cosmological systems.
\end{abstract}



\maketitle

\section{Introduction}

Recent work by the Bicep 2 Collaboration \cite{Ade:2014xna}, has resulted in increased interest in the existence of primordial gravitational waves.  Although these results where later disproven~\cite{bicep2,planck}, speculation about the characteristics of primordial gravitational waves is still very much alive~\cite{Gogoberidze:2007an,Kahniashvili:2005mp,Kahniashvili:2005qi,Kahniashvili:2006hy,Kahniashvili:2006zs,Kahniashvili:2008pf,Kahniashvili:2008pe,Kahniashvili:2009mf,Kisslinger:2004uc}.  Our ultimate goal is to determine the characteristics of these waves, given different cosmological theories, and if possible, find ways to observe them.  However, before we can do that we must first determine how to efficiently extract gravitational waves produced by turbulence in the relativistic plasma which dominated the universe shortly after the electroweak phase transition.  Can these waves be accurately calculated using linearilized approximations of gravity?  Does the chaotic nature of the General Relativity's nonlinearities significantly affect the solution~\cite{Hobill}?  Can an averaging process be used to quickly extract gravitational waves from a dynamic spacetime~\cite{Ellis}?

Primordial plasma turbulence is believed to have occurred as a result of stirring caused by bubble wall collisions and other chaotic events during inflation and the Electroweak phase transition.  The characteristic velocity of the turbulent eddies was calculated to be as high as 0.65~\cite{Kahniashvili:2008pf,Kahniashvili:2009mf}.  The magnetic fields around this time may have been as high as $10^{20}$ G.  This number was determined by extrapolating the accepted upper limit on cosmic magnetic fields, $10^{-9}$ G to a much smaller scale factor and assuming that the $B \propto a^{2}$ relationship holds~\cite{Durrer:2013pga}.

The author's previous work \cite{Garrison:2012ex} showed how the framework of numerical relativity could be used to study the cosmology of the early universe.  In this article, we expand on the techniques presented.  We utilize two General Relativistic Magnetohydrodynamic (GRMHD) codes in order to simulate a turbulent plasma similar to that of the universe when it was less than a second old.  One of these codes utilizes full general relativity modeled using a modified version of the BSSN equations in order to simulate the dynamical spacetime background while the other relies on a linearized approximation of gravity.  We characterize the gravitational waves produced and then compare the results produced by both codes.

Both the nonlinear and linear GRMHD codes have advantages.  By not solving Einstein's Equations, the linear code is much faster and requires less memory to run.  This is because the linear code does not have to solve nearly as many evolution equations.  Solving Einstein's Equations using a first order BSSN formalism adds about 35 additional evolution equations to the computer's workload.  However, utilizing a nonlinear GRMHD code can have advantages as well.  For example, the nonlinear code allows for nonlinear and dynamic spacetime solutions to evolve from initial conditions.  Also, the nonlinear code allows for the injection of gravitational waves into the initial data which would then effect the dynamics of the plasma field and evolution of the system~\cite{duez2}.  Because of this, it is important that we develop both codes and determine their accuracy and limitations.

\section{Overview of the Software}

Our code uses SI units for input and output and geometerized units, $c = \hbar = G = 1$, are used for all calculations within the code.  Throughout this article, we will refer to time and distance in units of meters.

As described in the preceding paper~\cite{Garrison:2012ex}, the codes used here were specifically developed to study early universe cosmology.  These codes are based on the Cactus Framework (www.cactuscode.org).  Cactus was originally developed to perform numerical relativistic simulations of colliding black holes but it's modular design has since allowed it to be used for a variety of Physics, Engineering and Computer Science applications.  It is currently being maintained by the Center for Computation and Technology at Louisiana Sate University.  Cactus codes are composed of a flesh (which provides the framework) and the thorns (which provide the physics).  The codes used within this work, FixedCosmo and SpecCosmo, are a collection of cactus thorns.  They are written in a combination of F90, C and C++.

Both codes utilize the relativistic MHD evolution equations proposed by Duez~\cite{duez1}.  Both codes are also designed to utilize a variety of different differencing schemes including 2nd order Finite Differencing, 4th order Finite Differencing and Spectral Methods.  The key difference between the codes is that while one is capable of solving Einstein's Equations directly (through a modified BSSN formulation) as well as the relativistic MHD equations, the other solves the relativistic MHD equations but simulates an expanding spacetime and estimates the gravitational wave background without solving directly Einstein's Equations.  This allows us to perform a test to determine under what conditions it is important to spend computational resources to solve Einstein's equations directly and if the gravitational waves are being correctly extracted from the nonlinear code.  The codes were thoroughly tested~\cite{Garrison:2012ex} and found to accurately model known GRMHD dynamics.  These tests included MHD waves induced by gravitational waves test, the consistency of cosmological expansion test and shock tests.

The initial data used was derived from work done by several projects involving primordial magnetic fields, phase transitions and early universe cosmology in general~\cite{Durrer:2013pga,islam,Kahniashvili:2008pf,Kahniashvili:2009mf,Kisslinger:2004uc}.  This study models an extremely high energy epoch of the universe shortly after the Electroweak phase transition when the universe was about $10^{-6}$ s old.  The authors chose this as the starting point for our study because it was the beginning of the Hadronic Epoch of the early universe.  It should be noted however, that the purpose of this paper is to test numerical techniques and not to exactly model the early universe.

\section{Evolution Equations}

The MHD equations used to evolve both numerical codes were based on Duez's evolution equations~\cite{duez1}.

\f
\partial_t \rho_* + \partial_j (\rho_* v^j) = 0 ,
\ff
\f
\partial_t \tilde{\tau} + \partial_i (\alpha^2 \sqrt{\gamma} ~T^{0i} - \rho_* v^i) = s ,
\ff
\f
\partial_t \tilde{S}_i + \partial_j (\alpha \sqrt{\gamma} ~ T^j_i) = \frac{1}{2} \alpha \sqrt{\gamma} ~ T^{ \alpha \beta} g_{\alpha \beta , i} ,
\ff
\f
\partial_t \tilde{B}^i  + \partial_j (v^j \tilde{B}^i - v^i \tilde{B}^j) = 0 .
\ff

The first equation comes from conservation of baryon number, the second derives from conservation of energy, the third is conservation of momentum and the fourth is the magnetic induction equation.  For this simulation we use Geodesic Slicing, $\alpha$ = 1.0, $\beta_i$ = 0.0 for both codes.  Here $\rho_*$ is conserved density, $v^j$ is velocity, $\tilde{\tau}$ is the energy variable, $\tilde{S}_i$ is the momentum variable, $s$ is the source term, $\alpha$ is the lapse term, $\gamma$ is the determinant of the three metric and $T^{ij}$ is the stress-energy tensor.  The tilde denotes that the term was calculated with respect to the conformal metric.  These variables are defined in terms of the stress tensor, primitive variables and gauge quantities below.

\f
\rho_* = \alpha \sqrt{\gamma} \rho u^0 ,
\ff
\f
\tilde{\tau} = \alpha^2 \sqrt{\gamma} T^{00} - \rho_*
\ff
\f
\tilde{S}_i = \alpha \sqrt{\gamma} T^0_i ,
\ff
\f
s = \alpha \sqrt{\gamma} [ (T^{00} \beta^i \beta^j + 2 T^{0 i} \beta^j + T^{ij}) K_{ij} - (T^{00} \beta^i + T^{0i}) \partial_i \alpha
\ff

The nonlinear code utilizes a first order version of the BSSN equations to simulate the background space-time.  For fixed gauge conditions, the modified BSSN equations as defined by Brown~\cite{Brown:2012me} are:
\begin{eqnarray}
\overline\partial_0 K &=&
 \alpha\left( \ta^{ij}\ta_{ij} + \frac{1}{3}\, K^2 \right) + 4 \pi \alpha ( \rho + S) \ .\label{eq:K-evol} \\
\overline\partial_0 \phi &=& -\frac{\alpha}{6}\, K  \ ,\label{eq:phi-evol}
\\
\overline\partial_0 \phi_i &=& -\frac{1}{6}\alpha \overline D_i K - \kappa^\phi {\cal C}_i \ ,
\\
\overline\partial_0 \cg_{ij} &=& -2\alpha\ta_{ij} \ ,\label{eq:gammatilde-evol}
\\
\overline\partial_0 \ta_{ij} &=& e^{-4\phi}\left[
 \alpha(\tR_{ij} - 8 \pi S_{ij})- 2\alpha \overline D_{(i}\phi_{j)} + 4\alpha \phi_i\phi_j 
+ \Delta\tilde\Gamma^k{}_{ij}(2\alpha\phi_k) \right]^{TF} 
\nonumber\\ 
 & & + \alpha K\ta_{ij} - 2\alpha\ta_{ik}\ta^k_{\; j} \ ,\label{eq:Atilde-evol}
\\
\overline\partial_0 \tilde\gamma_{kij} &=& -2\alpha\overline D_k\ta_{ij}
 - \kappa^\gamma  {\cal D}_{kij} \ ,\label{eq:dkijtilde-evol}
\\
\overline\partial_0 \tilde\Lambda^i &=& - \frac{4}{3}\alpha \tilde D^i K
 + 2\alpha\left( \Delta\tg^i{}_{k\ell}\ta^{k\ell} + 6\ta^{ij} \phi_j - 8 \pi  \cg^{ij} S_j \right) \ .
\label{eq:gamma-evol}
\end{eqnarray}

The bar denotes a derivative taken with respect to the fiducial metric (defined here to have a determinant of one) and the tilde again denotes a derivative taken with respect to the conformal metric.  Also, ${\cal C}_i$ and ${\cal D}_{kij}$ are constraint equations and $\kappa^\phi$ and $\kappa^\gamma$ are proportionality constants.  $\rho$, $S$, $S_j$ and $S_{ij}$ are source terms as found in the standard version of the BSSN equations.  Brown et al also defined:
\begin{eqnarray}
{\cal C}_i &=& \phi_i - \overline D_{i}\phi = 0 , \\
{\cal D}_{kij} &=&  \tilde{\gamma}_{k i j} - \overline{D}_k \tilde{\gamma}_{ij} = 0 , \\
\Delta\tilde{\Gamma}^i{}_{k\ell} &=& \frac{1}{2}\tilde{\gamma}^{ij}\left(
 \tilde{\gamma}_{k\ell j} + \tilde{\gamma}_{\ell kj} - \tilde{\gamma}_{jk\ell} \right) \ , \nonumber\\
    \tR_{ij}  & = & -\frac{1}{2} \tilde\gamma^{k\ell} \overline D_k \tilde\gamma_{\ell ij} 
      + \tilde\gamma_{k(i} \overline D_{j)} \tilde\Lambda^k 
		+ \tilde\gamma^{\ell m}\Delta\tilde\Gamma^k_{\ell m} \Delta\tilde\Gamma_{(ij)k} \\
		& & + \tilde\gamma^{k\ell} [ 2\Delta\tilde\Gamma^m{}_{k(i} \Delta\tilde\Gamma_{j)m\ell} 
			+ \Delta\tilde\Gamma^m{}_{ik} \Delta\tilde\Gamma_{mj\ell} ] 
		  \ ,
\label{eq:constraints}
\end{eqnarray}

These equations allow gravitational waves to appear organically from the turbulent plasma field.  For the linear code, we approximate the effect of an expanding spacetime and determine the gravitational wave spectrum without utilizing full general relativity.  We do this by solving the Friedmann Equations and the linearized gravitational wave equations~\cite{islam}. 
\begin{eqnarray}
\partial_{0} a &=& a H \\
\partial_{0} H  &=& -H^2 - \frac{4}{3} \pi (\rho + 3 p) \\
\partial_{0} \tilde{h}_{ij} &=& q_{ij} \\
\partial_{0} q_{ij} &=&  \nabla^{2}  \tilde{h}_{ij}  / a^2  - 2 (2 H^2 + \partial_{0} H) (\delta_{ij} \tilde{h}^{ij} - \tilde{h}_{ij}) + 16 \pi T_{ij}.
\end{eqnarray}

Where $\tilde{h}_{ij}$ are gravitational perturbations calculated by the linearized code and $q_{ij}$ are their time derivatives. 

\section{Extraction Methods}

The purpose of this paper is to determine if an averaging method is effective in extracting gravitational waves from a cosmological spacetime.  In this section, we will describe both averaging and Newman-Penrose, particularly as they apply to cosmological spacetimes.  As described by Ellis~\cite{Ellis}, the averaging method assumes that the spacetime can be written in terms of perturbations as shown below:
\begin{eqnarray}
\bar{g}_{ij} &\equiv& \langle {g}_{ij}  \rangle \\
\bar{g}_{ab} \bar{g}^{bc} &=& \delta^a_c \\
g_{ab} &=& \bar{g}_{ab} + \delta g_{ab} \\
g^{ab} &=& \bar{g}^{ab} + h^{ab} \\
g_{ab} g^{bc} &=& \delta^a_c.
\end{eqnarray}

Unlike linearized gravity, $\bar{g}^{ij} \neq \langle {g}^{ij}  \rangle$ and $h^{ab} \neq \delta g^{ab} \equiv g^{ae} g^{bf} (\delta g_{ef})$.  This leads to the following equations for the Christoffel symbols and curvature tensors.
\begin{eqnarray}
\Gamma^a_{bc} &=& \bar{\Gamma}^a_{bc} + \delta \Gamma^a_{bc} \\
R_{ab} &=& \bar{R}_{ab} + \delta R_{ab} \\
G_{ab} &=& \bar{G}_{ab} + \delta G_{ab}
\end{eqnarray}

For a cosmological spacetime, the bar terms are analogous to the unperturbed FRW spacetime while the delta terms correspond to the gravitational waves.  This assumes that the average of the perturbations is effectively zero.  Because of this, we can assume that $\delta g_{ij} = a^2 h_{ij}$, therefore
\f
g_{ij} = \bar{g}_{ij} + a^2 h_{ij}.
\ff

Once these perturbations are calculated from the spacetime using the averaging method, we still need to derive the gravitational wave spectrum, taking into account that the resulting gravitational perturbations are not necessarily transverse-traceless.  This will be discussed in the next section.  An obvious limitation of the averaging method is that it is not practical when the background spacetime contains singularities.  The Newman-Penrose formalism lacks such limitations but is much more complicated to utilize for a cosmological spacetime.

The Newman-Penrose formalism works by contracting the Weyl tensor ($C_{abcd}$) with a complex null tetrad in order to form 5 complex scalars, $\Psi_0$ ... $\Psi_4$.  The Weyl tensor is by definition the traceless part of the Riemann tensor and the tetrad is formed by a combination of real ingoing and outgoing vectors as well as a complex vector and its complex conjugate. In order to extract gravitational waves using the Newman-Penrose formalism, we would first need to calculate the 10 traceless components of the Riemann tensor and then determine the ingoing and outgoing vectors with relationship to the source. Once these quantities are known, gravitational waves can then derived from the $\Psi_0$ and $\Psi_4$ Weyl scalars, defined as
\begin{eqnarray}
\Psi_0 &=& -C_{abcd} l^a m^b l^c m^d \\
\Psi_4 &=& -C_{abcd} k^a \bar{m}^b k^c \bar{m}^d.
\end{eqnarray}

Here $l^a$ and $k^a$ are the outgoing and ingoing null vectors respectively, $m^b$ is the complex vector constructed by two spatial vectors that are orthogonal to the null vectors and $\bar{m}^b$ is the complex conjugate of $m^b$ .  Once calculated, $\Psi_4$ can be used to calculate gravitational wave luminosity and energy/angular momentum loss due to an outgoing gravitational wave.  The $\Psi_0$ Weyl scalar can be used to do the same for ingoing gravitational waves.  This brings up a major problem with using Weyl scalars to extract gravitational waves from a cosmological spacetime.  How does one define ingoing and outgoing gravitational waves when the source is everywhere?   Even if it was possible to define an appropriate complex null tetrad, wouldn't the computation needed to derive $\Psi_0$ and $\Psi_4$ Weyl scalars far exceed what is needed to extract gravitational waves using an averaging scheme?  Because of these questions, we believe that it was necessary to determine if averaging could be used for extracting gravitational waves from cosmological systems.

\section{Calculation of the Gravitational Wave Spectrum and Polarization}

The Gravitational Wave Spectrum is output as characteristic strain, energy density and degree of polarization by the numerical code \cite{moore}.  For the full GR code, we utilize an averaging process to calculate the gravitational waves produced by the simulation~\cite{Ellis}.  Here gravitation perturbations can be found by subtracting the mean value of the metric, $\bar{g}_{ij}$ from it's value at any point and correcting for the scale factor.

\f
h_{ij} = (g_{ij} - \bar{g}_{ij})/a^2
\ff

Unfortunately these perturbations are not necessarily in the transverse-traceless gauge so the energy density must be calculated as

\begin{eqnarray}
t_{00} &=& \frac{1}{32 \pi G} \langle (\partial_{0} h_{ij}) ( \partial_{0} h^{ij}) - \frac{1}{2} (\partial_{0} h) (\partial_{0} h) - (\partial_{\rho} h^{\rho \sigma})(\partial_{0} h_{0 \sigma}) - (\partial_{\rho} h^{\rho \sigma})(\partial_{0} h_{0 \sigma}) \rangle.
\end{eqnarray}

The brackets denote the average over several wavelengths.  By utilizing Geodesic Slicing conditions the last two terms are identically zero.  The time derivatives of the perturbations can be rewritten in terms of the extrinsic curvature (K), lapse ($\alpha$), Hubble Parameter ($H=\frac{\dot{a}}{a}$), and scale factor (a).

\f
\partial_{0} h_{ij}  \partial_{0} h^{ij} - \frac{1}{2} \partial_{0} h \partial_{0} h = 2 [( 2\alpha^2 K_{ij} K^{ij} + 4 \alpha H K + 6 H^2) -  (\alpha K + 3 H)^2] / a^4
\ff

The mass density is then normalized by dividing by the critical density

\f
\rho_c = \frac{3 H^2} {8 \pi G}.
\ff

This results in a normalized energy density,

\f
\Omega_{N} = -\langle (2 \alpha^2 K_{ij} K^{ij} + 4 \alpha H K + 6 H^2 - (\alpha K + 3 H)^2) / (6 a^4 H^2) \rangle.
\ff 

The normalized energy density for the linearized code is then simply calculated as

\f
\Omega_{L}=  \frac{1}{12 H^2 a^4} \langle q_{ij} q^{ij} - \frac{1}{2} (q_i^i)^2 \rangle.
\ff

For both codes, the magnitude of the characteristic strain is calculated using the quadratic mean of the plus and cross polarizations of the gravitational waves perceived to be traveling along the x, y and z axes.  The plus polarizations are calculated by finding the difference between the diagonal transverse perturbation terms while the cross terms relate directly to the perturbations in the off-diagonal transverse terms.
\begin{eqnarray}
h^{+}_{x}  &=&  \frac{1}{2} (h_{yy} - h_{zz}) \\
h^{+}_{y}  &=&  \frac{1}{2} (h_{zz} - h_{xx}) \\
h^{+}_{z}  &=&  \frac{1}{2} (h_{xx} - h_{yy}) \\
h^{\times}_{x}  &=& h_{yz} \\
h^{\times}_{y}  &=& h_{xz} \\
h^{\times}_{z}  &=& h_{xy}
\end{eqnarray}

\f
h_{i}(k,t) = \sqrt {| h^{+}_{i}(\vec{k},t) |^2 + | h^{\times}_{i}(\vec{k},t) |^2} 
\ff

\f
h(k,t) = \sqrt { h_{x}(k,t)^2 + h_{y}(k,t)^2 + h_{z}(k,t)^2}
\ff

According to Kahniashvili \cite{Kahniashvili:2006zs}, the degree of polarization can be defined as,

\f
P(k,t) = \frac{\langle h^{R*} (\vec{k},t) h^{R} ( \vec{k '} ,t) - h^{L*} (\vec{k},t) h^{L} (\vec{k '} ,t) \rangle} {\langle h^{R*} (\vec{k},t) h^{R} (\vec{k '} ,t) + h^{L*}  (\vec{k},t) h^{L} (\vec{k '} ,t) \rangle} 
\ff

Where the right and left polarizations for waves traveling along the x, y or z axis are defined as,
\begin{eqnarray}
h^{R}_i  =  h^{+}_i + i h^{\times}_i \\
h^{L}_i  = h^{+}_i - i h^{\times}_i
\end{eqnarray}

By expanding the left and right polarizations into plus and cross polarizations, the polarization degree can be rewritten as,

\f
P_{i} (k_{i},t) = \frac{2 \langle Im[h^{+}_{i}(k_{i},t)] Re[h^{\times}_{i}(k_{i},t)] - Re[h^{+}_{i}(k_{i},t)] Im[h^{\times}_{i}(k_{i},t)] \rangle} {\langle  |h^{+}_{i}(k_{i},t) |^2 + | h^{\times}_{i}(k_{i},t) |^2 \rangle} 
\ff

\section{Experimental Set-up}

Our simulation begins shortly after the electroweak energy scale, at the beginning of the Hadronic Epoch.  The numerical values for the initial conditions where calculated using the available literature~\cite{islam} and are the exact same for both codes.  These calculations give us an initial temperature of $1.30 \times 10^{13}$ K at time $1.0 \times 10^{-6}$ s.  The scale factor and Hubble Parameter are $a = 2.096 \times 10^{-13}$ and $H = 7.99 \times 10^{6} s^{-1}$ respectively.  The critical mass/energy density at the time was $1.14 \times 10^{23} \frac{kg}{m^3}$.  The characteristic velocity of the turbulent eddies for all runs was set to 0.25~\cite{Kahniashvili:2008pf,Kahniashvili:2009mf}.  The magnetic field at the time could have been as large as $10^{17}$ G so we set the amplitude of the magnetic fields to $10^{14}$ G for our simulations.

We ran three sets of simulations using the nonlinear and linear codes with the exact same initial conditions.  The first simulation utilized a grid 1000 m cubed with $64^3$ grid points and a timestep of $10^{-6}$ m. These dimensions were chosen so that any resulting gravitational waves would correspond to the frequency range of Pulsar Timing observations once universal expansion is taken into account.  The other two used higher or lower resolutions in order to establish convergence in our results.  Geodesic slicing conditions, periodic boundary conditions and Fourier spectral differencing were used for all simulation runs.  There were no shocks or discontinuities in the system so we did not utilize our High Resolution Shock Capturing (HRSC) routines.  We also used a 3rd order Iterative Crank Nicolson time scheme for time integration.  All runs started with no initial gravitational waves but the density, temperature, velocity and magnetic fields were all perturbed to model that of an early universe space-time~\cite{planck2,Durrer:2013pga,hall,huffenberger,kurki,mukherjee}.  The initial density and temperature were perturbed by random phase cosine functions with an amplitude proportional to their wavenumber, $k$, effectively a Fourier series.  The initial magnetic field consisted of random phase cosine waves with an amplitude proportional to their wavenumber squared, $k^2$.  The initial velocity field consisted of random phase cosine waves with an amplitude proportional to their wavenumber cubed, $k^3$.  Each run utilized 64 processing cores on the Maxwell computing cluster at the University of Houston's Center for Advanced Computing and Data Systems.  Over 10,000 iterations were produced for each data run.

\section{Results}

It should be noted that these results have not been extrapolated for present day observations and only represent a relatively short period in the early universe.  This is because of limits in our available computing resources to perform such a long simulation.  The strain and normalized energy density outputs for all runs appeared to be independent of k, so we chose to focus on the mean value of these quantities.  As one can see from Figure 1, the difference between the average strain as calculated by the nonlinear and linear codes is negligible.  We see that the same is true for the Energy Density as shown in Figure 2.  Based on the data, we can assume that strain and energy density calculations derived from the linear code is accurate to within a part in a thousand of those derived from the nonlinear code using the averaging technique.  

\begin{figure}  
\begin{center}
\includegraphics[height=3.5in,width=4.50in,angle=0]{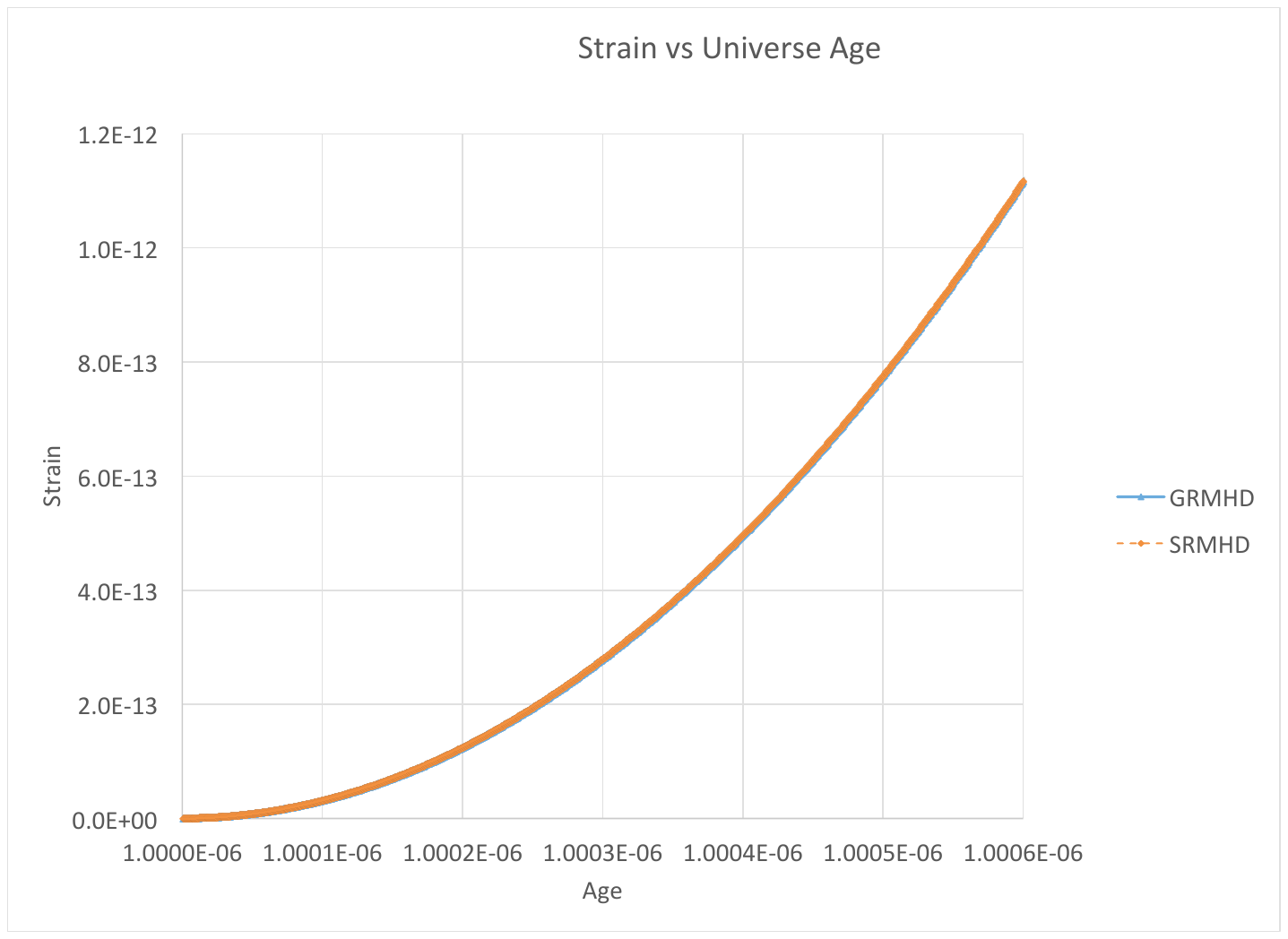}
\includegraphics[height=3.5in,width=4.50in,angle=0]{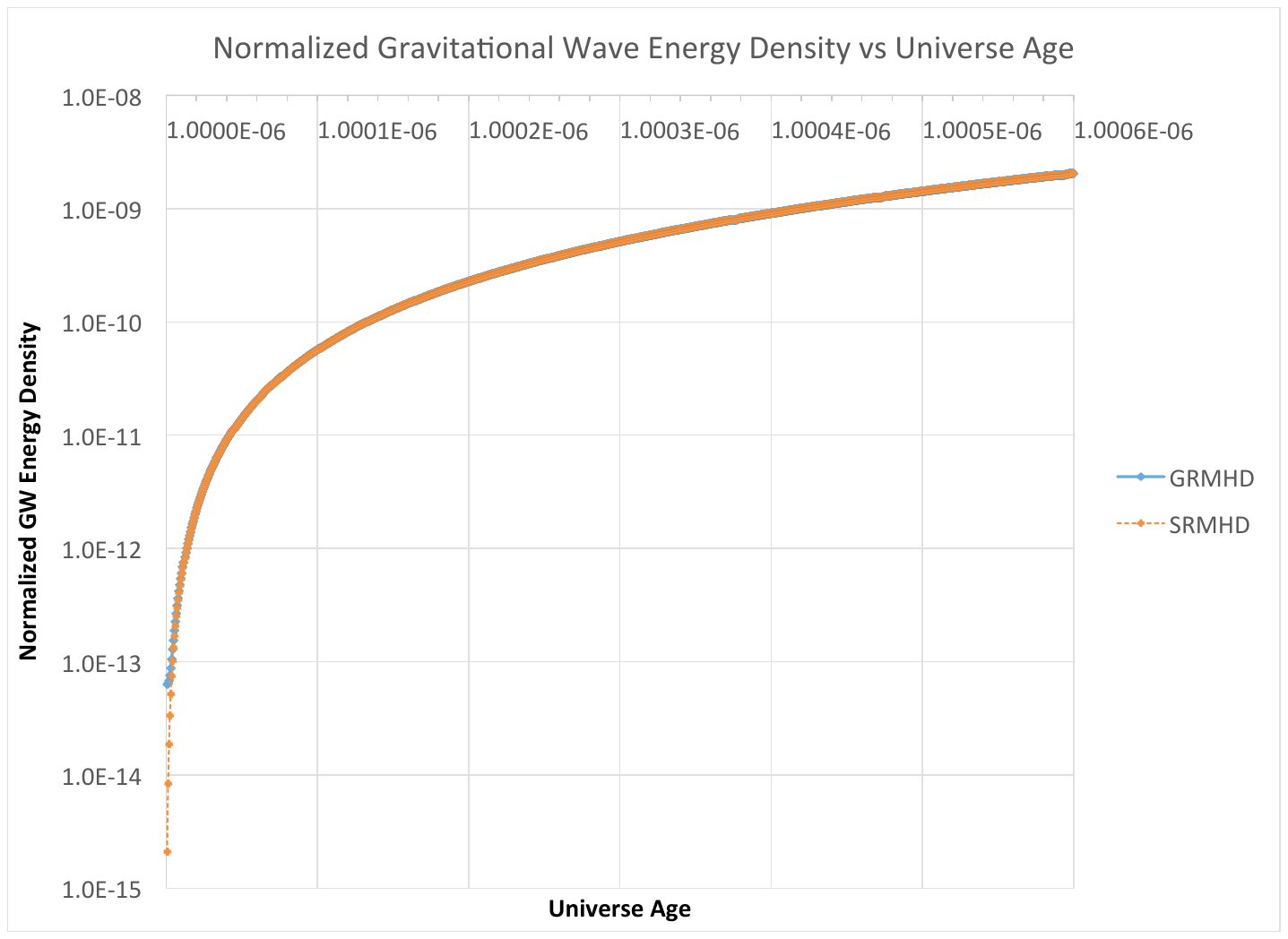}
\caption{\small \sl Strain of gravitational waves produced by the linear and nonlinear codes. Normalized Energy Density of gravitational waves produced by the linear and nonlinear codes.}  
\end{center}
\end{figure} 

\begin{figure}  
\begin{center}
\includegraphics[height=3.50in,width=4.50in,angle=0]{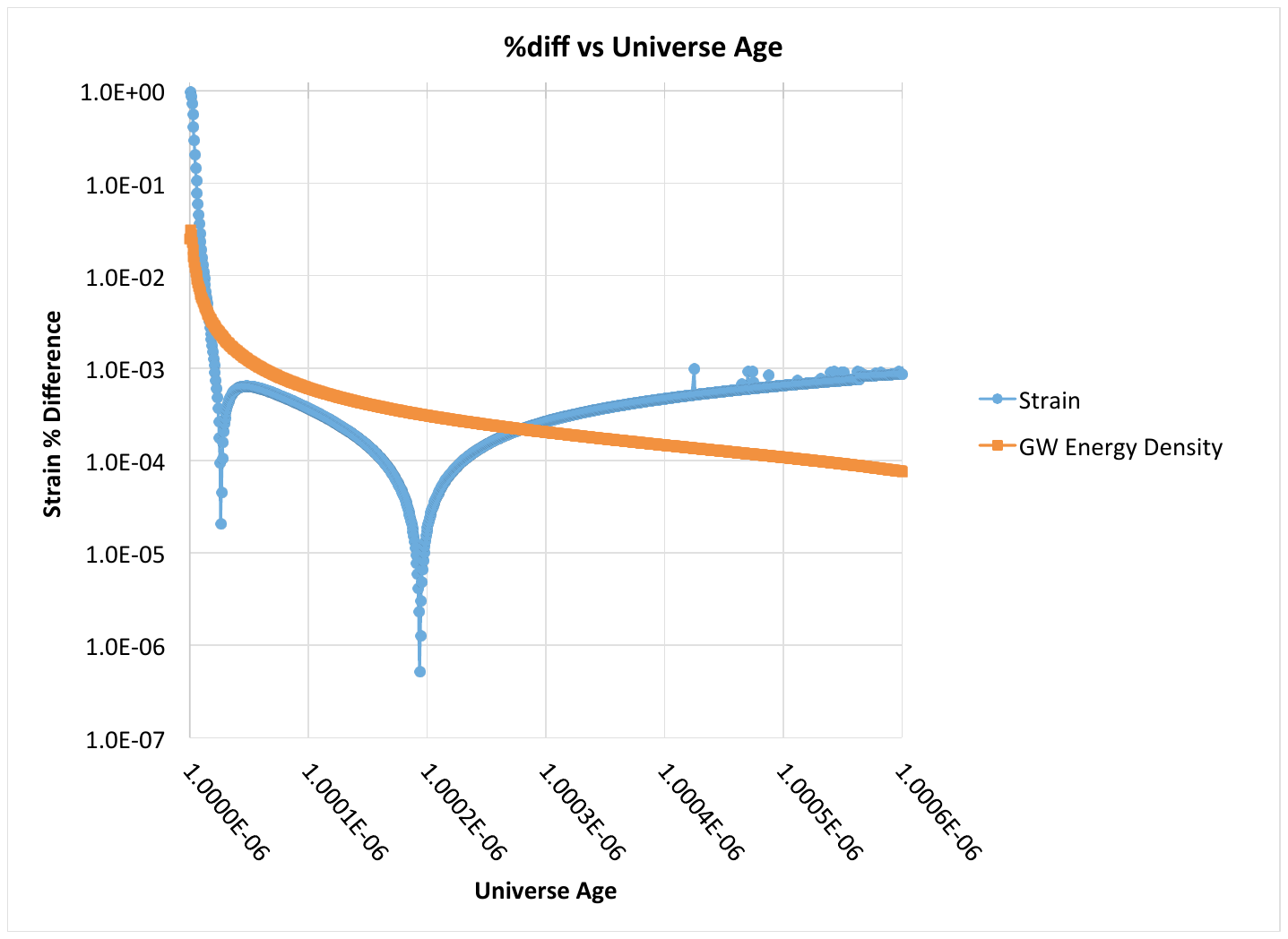}
\caption{\small \sl The relative error was calculated by dividing the difference in strain (energy density) by the strain (energy density) of the nonlinear code.}  
\end{center}
\end{figure} 

As shown in Table 1, The degree of polarization for the gravitational waves appears to be the same for both the linear and nonlinear calculations. These results appear to agree with work done by Kahniashvili~\cite{Kahniashvili:2005qi,Kahniashvili:2006zs}.  This calculation was performed from a sample of 5 different times between $1.0005 \times 10^{-6}$s and $1.0006 \times 10^{-6}$s.  As expected, since the initial data was isotropic, there does not appear to be any bias in the degrees of polarizations for the X, Y and Z directions or between left and right polarized gravitational waves generated by the turbulent system.  Further, the statistics of the polarizations for both the nonlinear and linear simulations appear to be the same to two significant figures.

\begin{table*}
\caption{Degree of Polarization at the final time. \label{table:Polarization}}
\centering
\begin{tabular}{c  c  c  c}
\hline \hline
 & X - Direction & Y - Direction & Z - Direction \\ 
\hline \hline
Nonlinear (Mean) &	0.0 &	 -1.56125E-17 & 0.0 \\
\hline
Linear (Mean) & 0.0 & 0.0 & 0.0 \\
\hline
Nonlinear (Standard Deviation) & 0.592164099 &	0.518579627 & 0.569916566  \\
\hline
Linear (Standard Deviation) & 0.592333947 & 0.515687302 & 0.572444906 \\
\hline \hline
\end{tabular}
\vskip 12pt
\label{tab:Polarization}
\end{table*} 

\section{Discussion}

In this article, we tested the averaging process to extract gravitational waves from nonlinear cosmological simulations.  We did this by using cosmological linear and nonlinear codes to solve the exact same problem and comparing the results.  We looked at three characteristics of the resulting gravitational waves, strain, normalized energy density and degree of polarization.  We found that the results agreed for all three measures.  The tiny differences that did occur may be partially explained by small nonlinear effects.  We believe that these results prove that the averaging process outlined in Section 4 of this paper is an accurate method of extracting gravitational waves for cosmological systems that lack spacetime singularities.  We have demonstrated that this holds true for gravitational waves with strains up to $10^{-12}$ and normalized energy densities as high as $10^{-8}$.  Further work is needed to test the limits of the averaging process.  Also, as a result of this work, we can conclude that the linear and nonlinear codes produce the same gravitational waves to within a part in $10^3$.  

\section{Acknowledgement}

The authors would like to acknowledge the support of the University of Houston Center for Advanced Computing and Data Systems for access to the high performance computing resources used for the completion of this project.  The authors would also like to thank Walter Thompson and Samina Masood for several useful conversations and helpful suggestions.


\end{document}